\documentclass[10pt,a4paper,twoside]{article}
\usepackage{epsfig}
\usepackage{baltlat5}
\usepackage{wrapfig}
\begin{document}
\ \ \vspace{-0.5mm}

\setcounter{page}{471}
\vspace{-2mm}

\titlehead{Baltic Astronomy, vol.\ts 15, 471--480, 2006.}

\titleb{A COMPARISON OF THE PUBLISHED STELLAR\\ PHOTOMETRY DATA IN THE
SOUTH-WEST FIELD OF\\ THE GALAXY M\,31 DISK}

\begin{authorl}
\authorb{D.~Narbutis}{1},
\authorb{R.~Stonkut\.{e}}{1,2} and
\authorb{V.~Vansevi\v{c}ius}{1}
\end{authorl}

\begin{addressl}
\addressb{1}{Institute of Physics, Savanori\c{u} 231, Vilnius
LT-02300, Lithuania \\ wladas@astro.lt}
\addressb{2}{Vilnius University Observatory, \v{C}iurlionio 29,
Vilnius LT-03100, Lithuania}
\end{addressl}

\submitb{Received 2006 September 22; accepted 2006 September 29}

\begin{summary} We compare stellar photometry data in the South-West
part of the M\,31 disk published by Magnier et al.  (1992), Mochejska et
al.  (2001) and Massey et al.  (2006) as the local photometric standards
for the calibration of star cluster aperture photometry.  Large
magnitude and color differences between these catalogs are found.  This
makes one to be cautious in using these data as the local photometric
standards for new photometry.  \end{summary}


\resthead{A comparison of the published photometry in M\,31}
{D.~Narbutis, R.~Stonkut\.{e}, V.~Vansevi\v{c}ius}

\sectionb{1}{INTRODUCTION}
The well calibrated local photometric standards established with small
to medium size telescopes are of great importance for accurate
photometry in the fields selected for study with the 8--10 meter class
telescopes.  In the course of compact star cluster study in the
South-West part of the M\,31 disk, Kodaira et al.  (2004) have used $B$,
$V$ and $R$ photometry from Magnier et al.  (1992, hereafter Mag92) for
calibration of their photometric data.  We compared the Mag92 $B$ and
$V$ dataset with the available stellar photometry data published by
Mochejska et al.  (2001; hereafter Moc01) and found significant
differences.  However, the Mag92 data were used by Kodaira et al.
because of availability of the $R$-band data.  Recently a new wide-field
photometric survey of M\,31 was conducted by Massey et al.  (2006;
hereafter Mas06).  The authors described their calibration procedures in
detail and compared their own photometry with Mag92 and Moc01 datasets
giving a hint of rather large discrepancies.

Narbutis et al.  (2006) used publicly
available\footnote{~http://www.lowell.edu/users/massey/lgsurvey.html}
mosaic images of the M\,31 galaxy from the Local Group Galaxies Survey
project performed by Mas06 to measure compact star clusters in the {\it
UBVRI} passbands.  For calibration of the cluster photometry we have
applied photometry datasets of Mag92, Moc01 and Mas06.  However, the
systematic differences of colors and magnitude zero-points, found by
Mas06 stimulated our efforts to study this problem in more detail.

\vskip2mm
\vbox{
\centerline{\psfig{figure=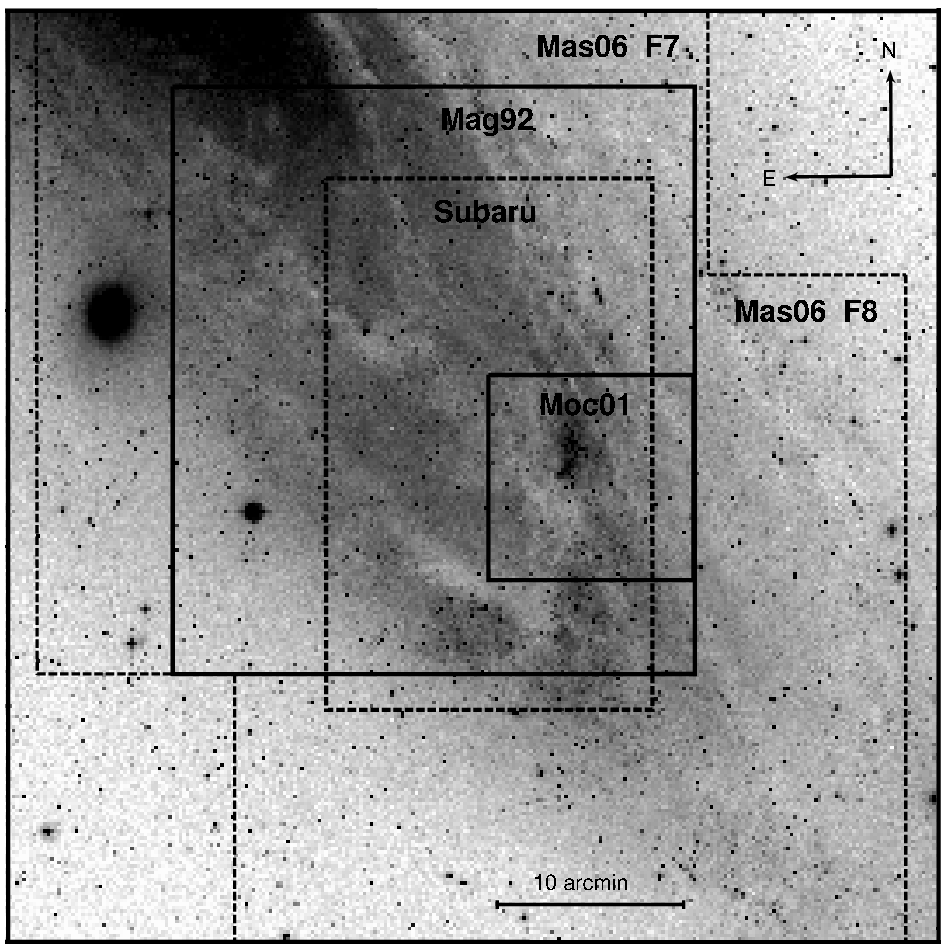,width=120truemm,angle=0,clip=}}
\vspace{1mm}
\captionb{1}{South-West area of the M\,31 galaxy disk (DSS $B$-band
image) with indicated fields under consideration.  ``Subaru'' marks the
field studied by Kodaira et al.  (2004) and Narbutis et al.  (2006).}}

\vskip2mm

\sectionb{2}{DATA}
The area selected from Mag92 for our investigation is bounded by the
following J2000.0 coordinates:  right ascension from 0$^{\rm
h}$\,39$^{\rm m}$\,57.5$^{\rm s}$ to 0$^{\rm h}$\,42$^{\rm
m}$\,26.5$^{\rm s}$ and declination from 40\degr\,32\arcmin\,24\arcsec\
to 41\degr\,04\arcmin\,12\arcsec\ ($\sim$28\arcmin$\times$32\arcmin),
covering parts of the fields F7 and F8 from Mas06, the area studied in
Kodaira et al.  (2004) and Narbutis et al.  (2006) and the field F from
Moc01 (see Figure 1 for area limits).

The Mag92 and Moc01 photometry data were taken from the
VizieR\footnote{~http://vizier.u-strasbg.fr/} catalog service
(Ochsenbein, Bauer \& Marcout 2000). The Mas06 catalog was taken from
their Table 4 (online version) published at the AJ Web site.

The typical seeing (full width at half maximum of stellar images, FWHM)
of the published photometric data are as follows:

Mag92 -- 2.3\arcsec\ (from 1.4\arcsec\ to 4.0\arcsec);

Moc01 -- 1.0\arcsec\ in $V$-band, 1.1\arcsec\ in $I$-band and
1.7\arcsec\ in $B$-band;

Mas06 -- 1.0\arcsec\ (from 0.8\arcsec\ to 1.4\arcsec).

The selection criteria of stellar objects from the compared catalogs
were as follows:

Mag92 -- photometric error $<$0.06 mag in $V$ band and $<$0.08 mag
in $B$ and $I$ bands, at least two observations in each passband
available and the object is classified as a star;

Moc01 -- photometric error $<$0.05 mag in each passband and
variability index $Js\,<\,1.5$ (Stetson 1996);

Mas06 -- photometric error $<$0.015 mag, at least four observations
in each passband available and faint magnitude limit $V=19.5$ mag.

Systematic object coordinate differences of Mag92, Moc01 and Mas06 data
sets exceeding 1\arcsec\ were found by examining regions, over-plotted
on the F7 field mosaic image from Mas06.  IRAF's (Tody 1993) $geoxytran$
procedure was used for nonlinear transformation of the object
coordinates given by Mag92 and Moc01 to the Mas06 field's F7 $V$-band
mosaic image coordinate system.  Finally, we achieved well matching
($<$0.2\arcsec) homogeneous coordinate systems of all three datasets and
cross-identified the stars.  Maximum object coordinate deviations of
0.5\arcsec\ and 0.7\arcsec\ were allowed for Moc01 and Mag92 star
identification with the Mas06 dataset, respectively.

The selection criteria and star coordinate matching limits applied
predetermined the final data sets used for further photometry comparison
consisting of:  343 and 336 stars from Mag92 possessing $B$--$V$ and
$V$--$I$ colors, respectively (Figure 2), and 233 and 213 stars from
Moc01 possessing $B$--$V$ and $V$--$I$ colors, respectively (Figure 3).
In the next section we show and briefly discuss the $B$-, $V$-, and
$I$-magnitude and color differences (Mag92 and Moc01 minus Mas06) vs.
corresponding magnitudes and colors from Mas06.

\sectionb{3}{RESULTS AND DISCUSSION}
Initially we tested the coordinate dependence of magnitude and color
differences of datasets Mag92 and Moc01 vs.  Mas06.  The Moc01
photometric data obtained with a single CCD and covering a rather small
field of 11\arcmin$\times$11\arcmin\ show no systematic differences
larger than 0.05 mag.  The Mag92 data were obtained with a CCD of a
small size field-of-view (6.5\arcmin$\times$5.7\arcmin), and our
comparison area (see Figure 1) contains $\sim$20 different Mag92 fields.
However, we notice only one break in $V-I$ color differences and it is
less than 0.1 mag.  Relatively small coordinate dependencies of
magnitude and color differences do not alter significantly the results
presented in Figures 2--7.

As it can be easily seen in Figures 2 and 3, $V$-magnitude data are well
calibrated and show no significant zero-point deviations.  However,
large systematic discrepancies of the Mag92 and Moc01 zero-points in
respect to Mas06 are obvious for $B$ and $I$ magnitudes.  This effect,
most probably, occurs because of $B$--$V$ and $V$--$I$ color reduction
inaccuracy due to considerable color equations of the instrumental
systems and narrow color range of the standard stars used.

The differences of magnitudes and colors of the Mas06 and Mag92 from one
side and of Moc01 from the other side (plotted vs.  $B$--$V$) are shown
in Figures 4 and 5. Corresponding plots vs.  $V$--$I$ are shown in
Figures 6 and 7. These plots

\vbox{
\centerline{\psfig{figure=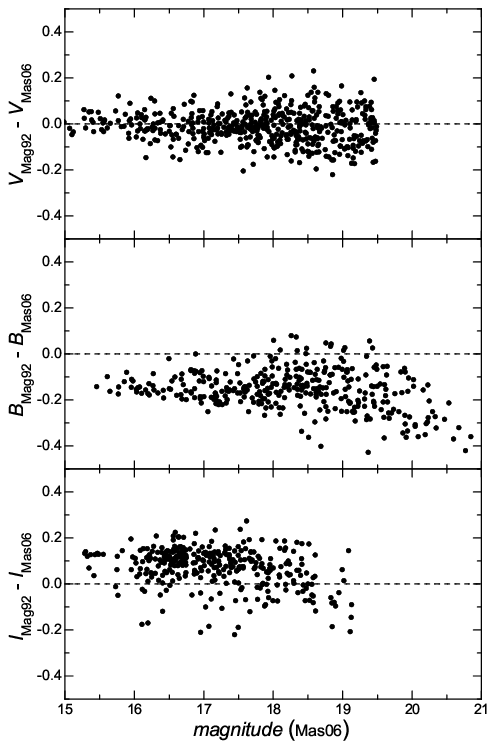,width=120truemm,angle=0,clip=}}
\vspace{-5mm} \captionb{2}{Differences of $B$, $V$ and $I$ magnitudes
from Mag92 and Mas06 plotted vs. corresponding magnitudes from Mas06.}}

\vbox{
\centerline{\psfig{figure=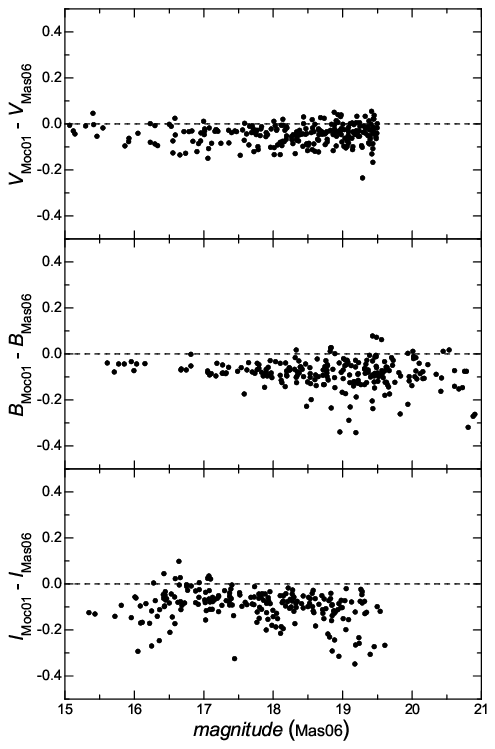,width=120truemm,angle=0,clip=}}
\vspace{-5mm} \captionb{3}{Differences of $B$, $V$ and $I$ magnitudes
from Moc01 and Mas06 plotted vs. corresponding magnitudes from Mas06.}}

\vbox{
\centerline{\psfig{figure=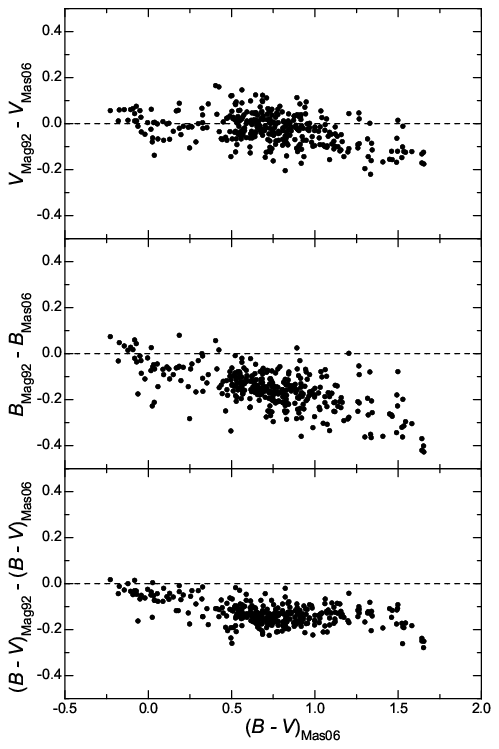,width=120truemm,angle=0,clip=}}
\vspace{-5mm} \captionb{4}{Differences of $B$ and $V$ magnitudes and
$B$--$V$ colors from Mag92 and Mas06 plotted vs.  $B$--$V$ from Mas06.}}

\vbox{
\centerline{\psfig{figure=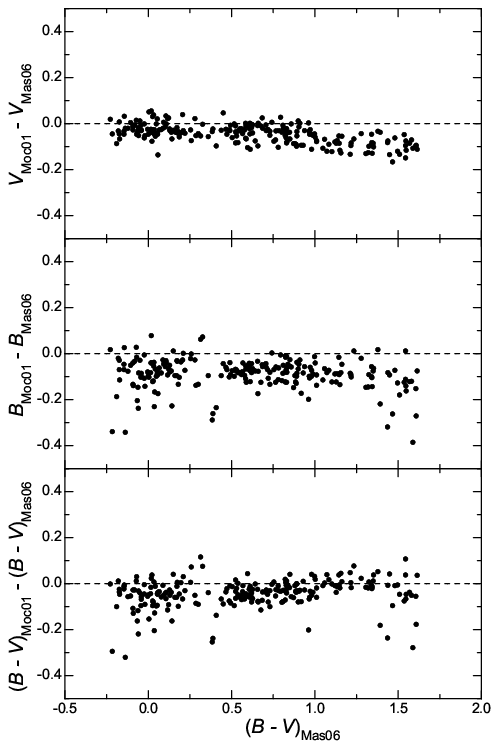,width=120truemm,angle=0,clip=}}
\vspace{-5mm} \captionb{5}{Differences of $B$ and $V$ magnitudes and
$B$--$V$ colors from Moc01 and Mas06 plotted vs.  $B$--$V$ from Mas06.}}

\vbox{
\centerline{\psfig{figure=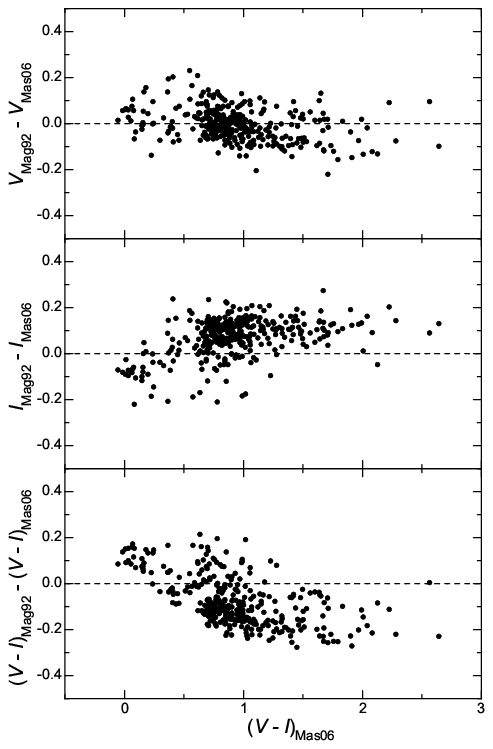,width=120truemm,angle=0,clip=}}
\vspace{-5mm} \captionb{6}{Differences of $V$ and $I$ magnitudes and
$V$--$I$ colors from Mag92 and Mas06 plotted vs.  $V$--$I$ from Mas06.}}

\vbox{
\centerline{\psfig{figure=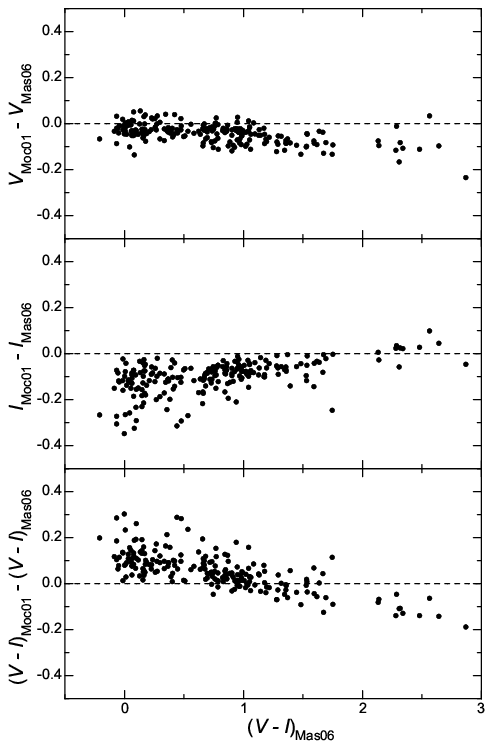,width=120truemm,angle=0,clip=}}
\vspace{-5mm} \captionb{7}{Differences of $V$ and $I$ magnitudes and
$V$--$I$ colors from Moc01 and Mas06 plotted vs.  $V$--$I$ from Mas06.}}

\newpage

\noindent confirm the above-stated presumption about color equation
determination difficulties and strongly imply that even carefully
calibrated CCD photometry cannot guarantee reliable photometric data
suitable, e.g., for star formation history analysis even in such a
well-studied galaxy as M\,31.

Kaluzny et al.  (1998) discussed the discrepancy and nonlinearity of
their $V$--$I$ colors in respect to Mag92 data.  We show that this
discrepancy is easily seen in Figure 6 as well, where Mag92 and Mas06
photometric data are compared.  However, $V$--$I$ colors of Moc01 also
show large systematic deviations from Mas06.  In both cases, huge color
deviations, as large as $\sim$0.3 mag, can hardly be accounted for color
equation, crowding and seeing or non-photometric weather effects.

Solely on the ground of Figures 2--7 it is rather difficult to decide
which of these three datasets is the most reliable.  The careful
reduction and calibration method used by Mas06, as well as internal
consistency check of overlapping fields, makes us believe that this is
the most accurately calibrated photometry dataset in the M\,31 galaxy
to date.

The surprisingly large magnitude and color differences between the
carefully calibrated Mas06, Mag92 and Moc01 catalogs, found in this
paper, suggest that users should be cautious using the published
photometric data as local photometric standards for calibration of the
new photometry.

\vskip5mm

ACKNOWLEDGMENTS.  This work was financially supported in part by a Grant
T-08/06 of the Lithuanian State Science and Studies Foundation.  This
research has made use of the SAOImage DS9 developed by Smithsonian
Astrophy\-si\-cal Observatory, and the Digitized Sky Surveys produced at
the Space Telescope Science Institute under U.S.  Government grant NAG
W-2166.

\vskip5mm

\References
\refb Kaluzny~J., Stanek~K.  Z., Krockenberger~M., Sasselov~D.  D.,
Tonry~J.  L., Mateo~M. 1998, AJ, 115, 1016

\refb Kodaira~K., Vansevi\v{c}ius~V., Brid\v{z}ius~A., Komiyama~Y.,
Miyazaki~S., Stonkut\.{e}~R., \v{S}ablevi\v{c}i\={u}t\.{e}~I.,
Narbutis~D. 2004, PASJ, 56, 1025

\refb Magnier~E.  A., Lewin~W.\,H.\,G., van Paradijs~J., Hasinger~G.,
Jain~A., Pietsch~W., Truemper~J. 1992, A\&AS, 96, 379

\refb Massey~P., Olsen~K.\,A.\,G., Hodge~P.  W., Strong~S.  B.,
Jacoby~G.  H., Schlingman~W., Smith~R.  C. 2006, AJ, 131, 2478

\refb Mochejska~B.  J., Kaluzny~J., Stanek~K.  Z., Sasselov~D.  D. 2001,
AJ, 122, 1383

\refb Narbutis~D., Vansevi\v{c}ius~V., Kodaira~K.,
\v{S}ablevi\v{c}i\={u}t\.{e}~I., Stonkut\.{e}~R., Brid\v{z}ius~A. 2006,
Baltic Astronomy, 15, 461 (this issue)

\refb Ochsenbein~F., Bauer~P., Marcout~J. 2000, A\&AS, 143, 221

\refb Stetson~P.  B. 1996, PASP, 108, 851

\refb Tody~D. 1993, in {\it Astronomical Data Analysis Software and
Systems II}, eds.  R. J.~Hanisch, R.\,J.\,V.~Brissenden \& J.~Barnes,
ASP Conf.  Ser., 52, 173

\end{document}